\pdfoutput=1
\documentclass[aps,prl,twocolumn,showpacs,superscriptaddress,groupeaddress,floatfix]{revtex4}  

\usepackage{graphicx}  
\usepackage{dcolumn}   
\usepackage{bm}        
\usepackage{amssymb}   
\usepackage[hyperindex,breaklinks]{hyperref}
\usepackage{multirow}
\usepackage{epsfig}
\usepackage{epstopdf}
\usepackage{amsmath,booktabs,array}
\usepackage{float}
\usepackage{eqparbox}
\usepackage{subfigure}
\usepackage{xspace}
\usepackage{slashed}

\def\ssWW{\ensuremath{ W^{\pm}W^{\pm}jj }\xspace}
\def\ts{\ensuremath{ \theta^{*} }\xspace}
\def\tsNN{\ensuremath{ \theta^{NN} }\xspace}
\def\cts{\ensuremath{ \cos\ts }\xspace}
\def\ctsNN{\ensuremath{ \cos\tsNN }\xspace}
\def\ctsnb{\ensuremath{ \cos \ts }\xspace}
\def\ctsSq{\ensuremath{ \cos^{2}\ts }\xspace}

\usepackage[percent]{overpic}
\usepackage{xcolor}

\begin{document}
\title{Determination of the $\boldsymbol{WW}$ polarization fractions in $\boldsymbol{pp \rightarrow W^\pm W^\pm jj}$ using a deep machine learning technique}
\author{Jacob Searcy}
\affiliation{Department of Physics, University of Michigan, Ann Arbor, Michigan 48109, USA}
\author{Lillian Huang}
\affiliation{Department of Physics, University of Michigan, Ann Arbor, Michigan 48109, USA}
\author{Marc-Andr\'e Pleier}
\affiliation{Brookhaven National Laboratory, Upton, New York 11973-5000, USA}
\author{Junjie Zhu}
\affiliation{Department of Physics, University of Michigan, Ann Arbor, Michigan 48109, USA}

\date{\today}

\begin{abstract}
The unitarization of the longitudinal vector boson scattering (VBS) cross section by the Higgs boson is 
a fundamental prediction of the Standard Model which has not been experimentally verified. One of the most
promising ways to measure VBS uses events containing two leptonically-decaying same-electric-charge 
$W$ bosons produced in association with two jets. However, the angular distributions of the leptons in the $W$ boson rest frame, which are commonly used to fit polarization fractions, are not readily available in this process due to the presence of two neutrinos in the final state. 
In this paper we present a method to alleviate this problem by using a deep machine learning technique to recover these angular distributions from measurable event kinematics and demonstrate how the longitudinal-longitudinal scattering fraction could be studied. We show that this method doubles the expected sensitivity when compared to previous proposals.

\end{abstract}

\maketitle

Studying longitudinal Vector Boson Scattering (VBS) processes has
long been a central goal of high energy
colliders~\cite{SSC_1}. Without a Higgs boson, the scattering
amplitude of longitudinal vector bosons ($V_{L}V_{L} \rightarrow
V_{L}V_{L}$) increases with center-of-mass energy and eventually
violates unitarity~\cite{Uni_1,Uni_2,Uni_3}. The discovery of a
Higgs-like boson at the LHC~\cite{ATLAS_higgs,CMS_higgs} was the first
step towards understanding these interactions. However, if this Higgs
boson's couplings to vector bosons deviate from the Standard Model
(SM) expectation the scattering amplitude of VBS processes can still increase with center-of-mass
energy, which makes VBS a sensitive probe of anomalous Higgs
couplings~\cite{Higgs_con}. In addition, many new physics scenarios
predict increases in
VBS cross sections, through
extended Higgs sectors or other new resonances~\cite{Tmatrix,VLVLBSM,Trip_Higgs_old,Trip_higgs_new}. VBS measurements hence
are both a window to new physics and a constraint on fundamental properties
of the Higgs boson. Measuring VBS processes at a hadron collider is
experimentally challenging due to small cross sections and the
difficulty of separating longitudinal states from transverse ones.

The ATLAS and CMS collaborations recently provided the first evidence
for and study of a VBS process using events with two leptonically
decaying same-electric-charge $W$ bosons in association with two
forward jets ($pp \to$ \ssWW)~\cite{ATLAS_ssWW,CMS_ssWW}. This final
state has the advantage of relatively small SM background
contributions compared to other VBS processes, paired with a
production rate large enough to measure in early LHC data sets.  While
an ideal candidate for first observation of the VBS process, measuring
the longitudinal fraction in these events is not straightforward since
the presence of two neutrinos in the final state prevents full
kinematic reconstruction of the events.

Recent studies have shown that advances in machine learning can
improve the prospects for measurements at the Large Hadron
Collider~\cite{Baldi:2014kfa,Baldi:2014pta}.  In this paper we explore
a machine learning technique that has not previously been used in the
experimental high energy physics community: regression with deep
neural networks. We apply this method to the difficult problem of
measuring longitudinal VBS in \ssWW.

In general, the polarization of a gauge boson can be determined from
the angular distribution of its decay products. The differential cross section of a
leptonically-decaying $W$ boson is related to the polarization fractions as~\cite{Wpol}:
\begin{multline}
 \label{eq:pol}
 \frac{d\sigma}{d\cts} \propto  \frac{3}{8} f_- (1 \mp \cts)^2 + \frac{3}{8} f_+ (1 \pm \cts)^2 \\ 
+\frac{3}{4} f_L (1 - \ctsSq) , {\text {for~}} W^\pm 
\end{multline}

where \ts is the angle between the charged lepton in the boson
rest frame and the $W$ boson direction of motion.  Fraction
parameters $f_{-}$, $f_{+}$ and $f_L$ denote the fractions of
events with the three possible polarization states of the $W$ boson, $-1$, $+1$ and 0,
respectively.  They are constrained via $f_- + f_+ + f_L = 1$.  In
order to measure \ts, we need to fully reconstruct the direction of
motion of the $W$ boson.

Requiring both $W$ bosons to decay leptonically in  $pp
\to$ \ssWW events enables the determination of the electric charge of
each $W$ boson via the charged leptons. However, since the
corresponding two neutrinos in the final state are not detected, the
$W$ boson rest frames cannot be directly measured.  It is thus
difficult to determine polarization fractions of each boson and
the fraction of longitudinal scattering events in the \ssWW process.

Many proposals have been made to determine the longitudinal fractions in other VBS final states, such as semi-leptonic
$W^+W^-$~\cite{Han:2009em} and $W^\pm Z$~\cite{VBSCuts1} or
fully-leptonic decay modes of $W^\pm Z$ and $ZZ$, where the full
event kinematics can be reconstructed or estimated using the $W$ boson mass constraint. 
However, these channels suffer from large SM backgrounds that are not present in the \ssWW channel. Attempts have been made 
to gain sensitivity through other variables~\cite{SSC_1,VBSCuts1,VBSME,Doroba:2012pd,new_VBS_warsaw_cut} than \ts in the \ssWW
channel. One example is the variable $R_{pT}=(p_{T}^{\ell 1} \times
p_{T}^{\ell 2}) / (p_T^{j_1} \times p_T^{j_2})$~\cite{Doroba:2012pd},
where $\ell_1$ and $\ell_2$ denote the two leptons in no particular
order and $j_1$ and $j_2$ denote the two most energetic jets in the
event. This single variable does not encompass all of the
sensitivity to longitudinal scattering, and better discrimination can
be achieved by combining the available event information with a
machine learning technique. Therefore, we develop a method to use a
neural network (NN) to map measurable quantities to the true \cts values
that contain the event's polarization information. This
approximation mitigates the limitation of missing kinematic information
in this final state, and makes \ssWW a promising channel for observing
the behavior of longitudinal $W$ boson scattering.

While it has become a common practice in high energy physics to use
multi-variate techniques to separate signal from background,
multi-variate regression is not commonly used to measure underlying
physics quantities of interest.  Unlike classification where the goal
of the neural network is to produce discrete assignments, for example
signal and background, NN regression relies on the fact that a neural
network is a universal approximator~\cite{NN_1}, to instead
approximate an unknown continuous function.  The goal of our NN is to
find the best approximation of the two truth values of \cts (one for
each $W$ boson) present in each event, using measurable
quantities. Similar techniques are currently in use to address the
problem of estimating parton distribution functions~\cite{NNPDF}.

It has also become common practice to train these multi-variate
methods using variables built from basic measurable event quantities
to be sensitive to a given physics process~\cite{Bhat:2010zz}. These high-level variables
add some understanding of the underlying physics to the training
variables and can produce better results. However, it has also
recently been observed that extending a single layer neural network to
a ``deep'' network containing many layers can regain most of the
sensitivity produced by these high-level
variables~\cite{Baldi:2014kfa,Baldi:2014pta}. Since only a few
high-level variables have been proposed for this process we choose to
use deep neural networks for maximum sensitivity.

\ssWW events have the signature of two quarks, two same sign leptons, and
two undetected neutrinos.  We use all basic measurable object
kinematics as input variables to the neural network: the transverse
momentum ($p_T$), pseudorapidity ($\eta$) and azimuthal angle ($\phi$)
of the two leptons and two jets, and $x$- and $y$-components of
missing transverse energy ($\slashed{E}_T^x$ and $\slashed{E}_T^y$).
The overall number of measurable quantities used hence is 14. The
goal of the multi-variate technique is to find the best mapping from
these measurable quantities to the two truth values of \cts (one for
each $W$ boson) present in each event.

Training deep NNs has been the subject of intensive study and a good
review of some of this work is presented in~\cite{NN_Review}. For
completeness we list some of the properties of the NN that we utilize.
We choose a multi-layer neural network with a two node final output
layer with linear activation to approximate the true $\cts$
distribution of each $W$ boson.  The NN is implemented with the Theano
software packages~\cite{theano1,theano2}. The cost function is defined
as
\begin{equation}
 \label{eq:cost}
{\cal{C}}= \frac{1}{N}\sum_{i=1}^{N} [(\ctsnb_{1,i} -\cos \theta^{NN}_{1,i})^2+(\ctsnb_{2,i}-\cos \theta^{NN}_{2,i})^2],
\end{equation}
where $N$ is the number of events per mini-batch, $\ctsnb_{1/2,i}$ is
the truth value of \cts for each $W$ boson with random ordering for
the $i$-th event, and $\cos \theta^{NN}_{1/2, i}$ is the value of the
two neural network outputs. A stochastic gradient descent algorithm is
used to train the neural network by minimizing the cost function.
Hyper-parameters are tuned by hand and confirmed by a local grid
search, with the best performance given by a deep network with 20
layers each with 200 hidden nodes.  This network yields a 20\% better
cost value in a validation data set than the best single layer NN
tested.

Signal \ssWW events are generated using the {\sc madgraph} event
generator~\cite{madgraph} at a proton-proton center-of-mass energy of
13 TeV.  The invariant mass of the two outgoing partons was required
to be greater than 150 GeV. The $W$ bosons are decayed, assuming they
are on-shell and have no spin correlations, with the {\sc decay}
routine provided with {\sc madgraph}. 500,000 events were generated
for training and another million for testing and validation.

Polarization fractions can then be obtained experimentally by fitting
the two-dimensional distribution of the NN output \ctsNN.  In order to
fit for these polarization fractions templates must be built for
``pure'' polarization states. These templates are created using
generator level helicity information. In addition, a method was tested
that used truth level reweighting and included all spin correlations
and off-shell effects, and the results were found to be comparable.

Since there are two $W$ bosons in each event, six distinct polarization states are possible. Events where both $W$ bosons have a polarization of 0 are referred to as $LL$, of 1 as $++$, and of $-1$ as $--$. Events with differing polarizations of ($-1$,1) or (1,$-1$) are referred to as $+-$, ($-1$,0) or (0,$-1$) as $L-$ and likewise (1,0) or (0,1) as $L+$. 

\begin{figure*}
\begin{tabular}{ccc}
\includegraphics[width=.32\textwidth,height=.18\textheight]{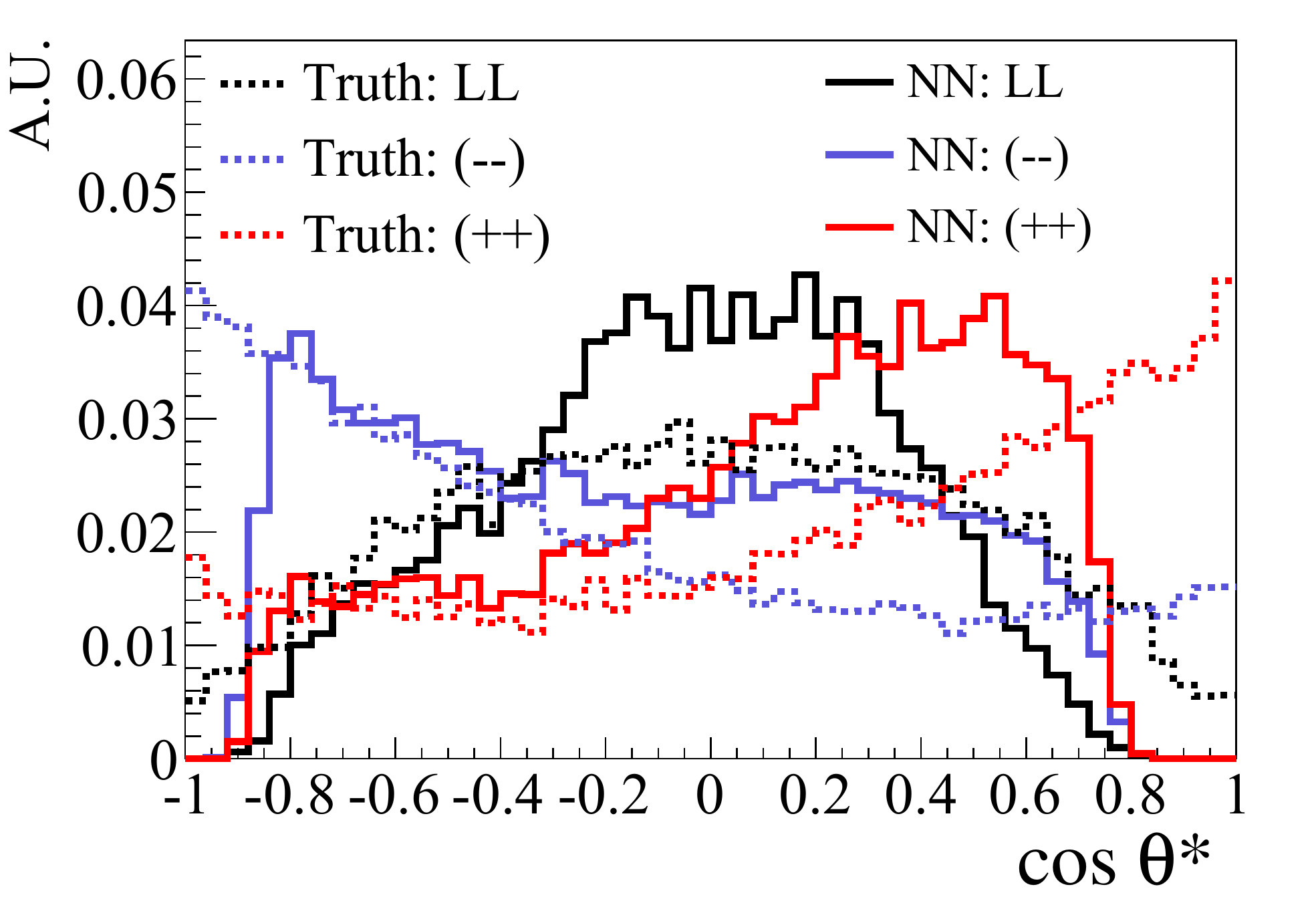}&
\includegraphics[width=.32\textwidth,height=.18\textheight]{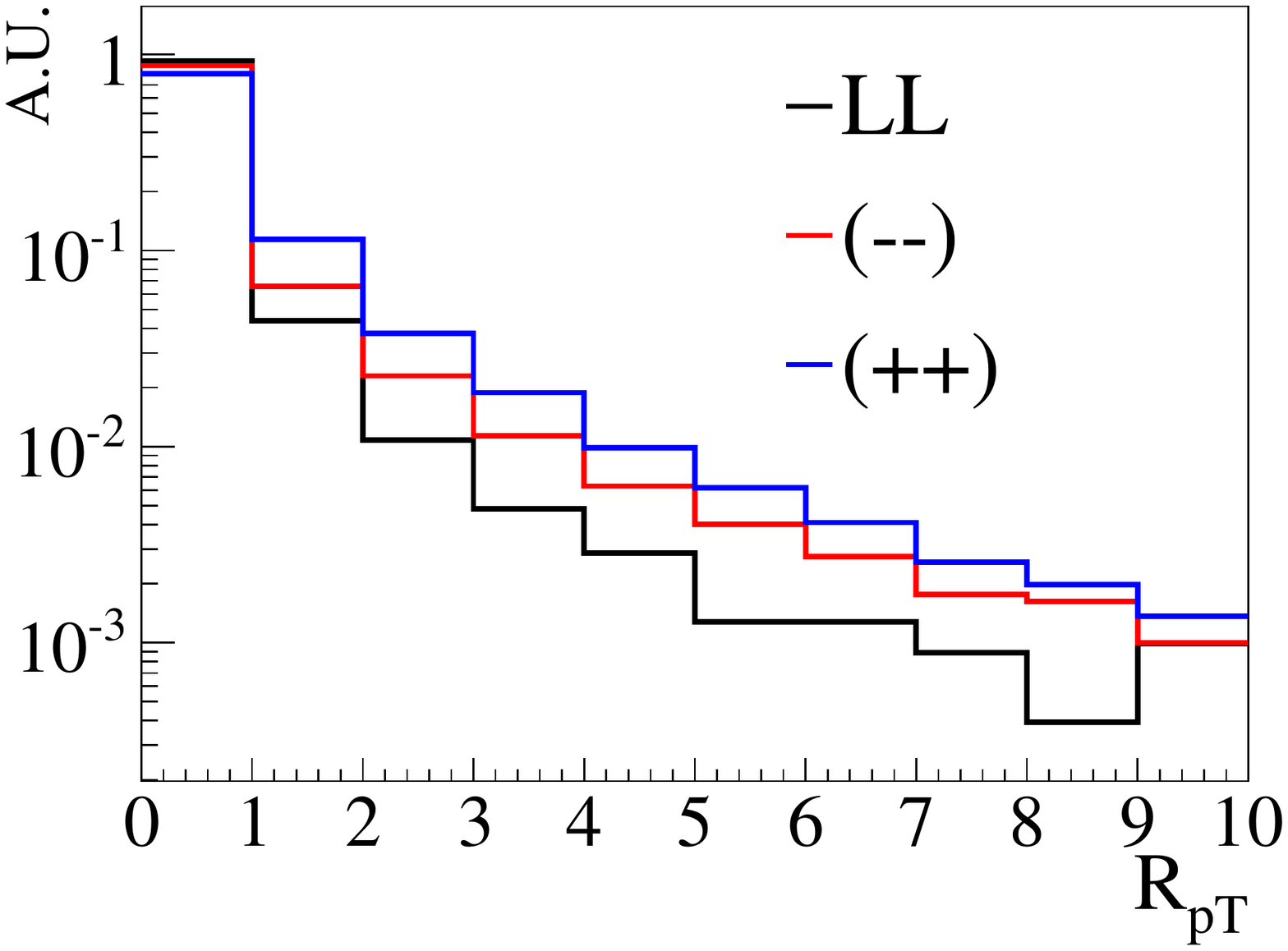}&
\includegraphics[width=.32\textwidth,height=.18\textheight]{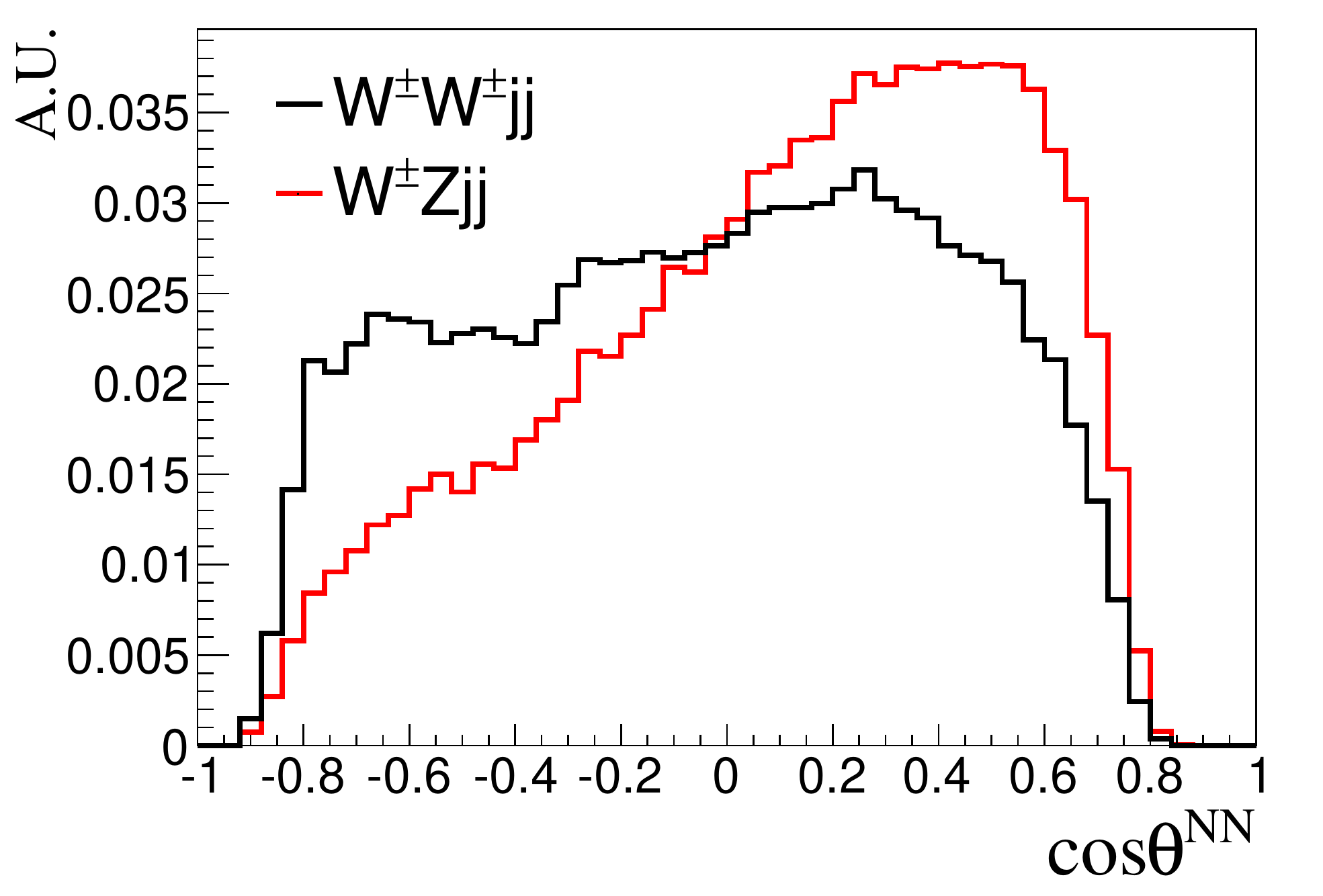}\\
(a)&(b)&(c)\\
\end{tabular}
\caption{\label{fig:polarization_comparison} A comparison of the truth
level \cts and the NN output \ctsNN distributions for $--$, $++$ and
$LL$ events (a), $R_{pT}$ templates for the corresponding polarization
states with log scaling (b), and comparison of the \ctsNN for the
signal and dominant $WZjj$ background (c) are shown with arbitrary
units (A.U.). (a) and (c) are the projections of the 2D distributions
onto one of the two leptons, which are identical to the projections
onto the other lepton due to arbitrary sorting.}
\end{figure*}

Figure~\ref{fig:polarization_comparison}(a) shows the comparison between the truth \cts and 
the NN output \ctsNN for $--$, $++$ and $LL$ events ( $+-/L+/L-$ are omitted for clarity, but closely resemble combinations of the templates shown). As expected, \ctsNN has less separation 
power for the different polarization states compared to \cts due to missing information for the two final state neutrinos. 
However, reasonable discrimination between each polarization state can clearly be seen from these distributions. 
Figure~\ref{fig:polarization_comparison}(b) shows the $R_{pT}$ distribution for $--$, $++$, and $LL$ events. 
The discrimination power is seen only for large values of $R_{pT}$, and only apparent with a logarithmic scale. Figure~\ref{fig:polarization_comparison}(c) shows the \ctsNN distribution for the signal \ssWW process and an important background process: $WZjj$ production.  Reasonable separation power is observed, which could be utilized in a combined fit. In an actual data analysis, the $WZjj$ component would be subtracted as background
from the observed data before fitting the polarization fractions.

\begin{figure*}
\begin{tabular}{ccc}
$TT$ &$TL$&$LL$\\
\includegraphics[width=.28\textwidth,height=.18\textheight]{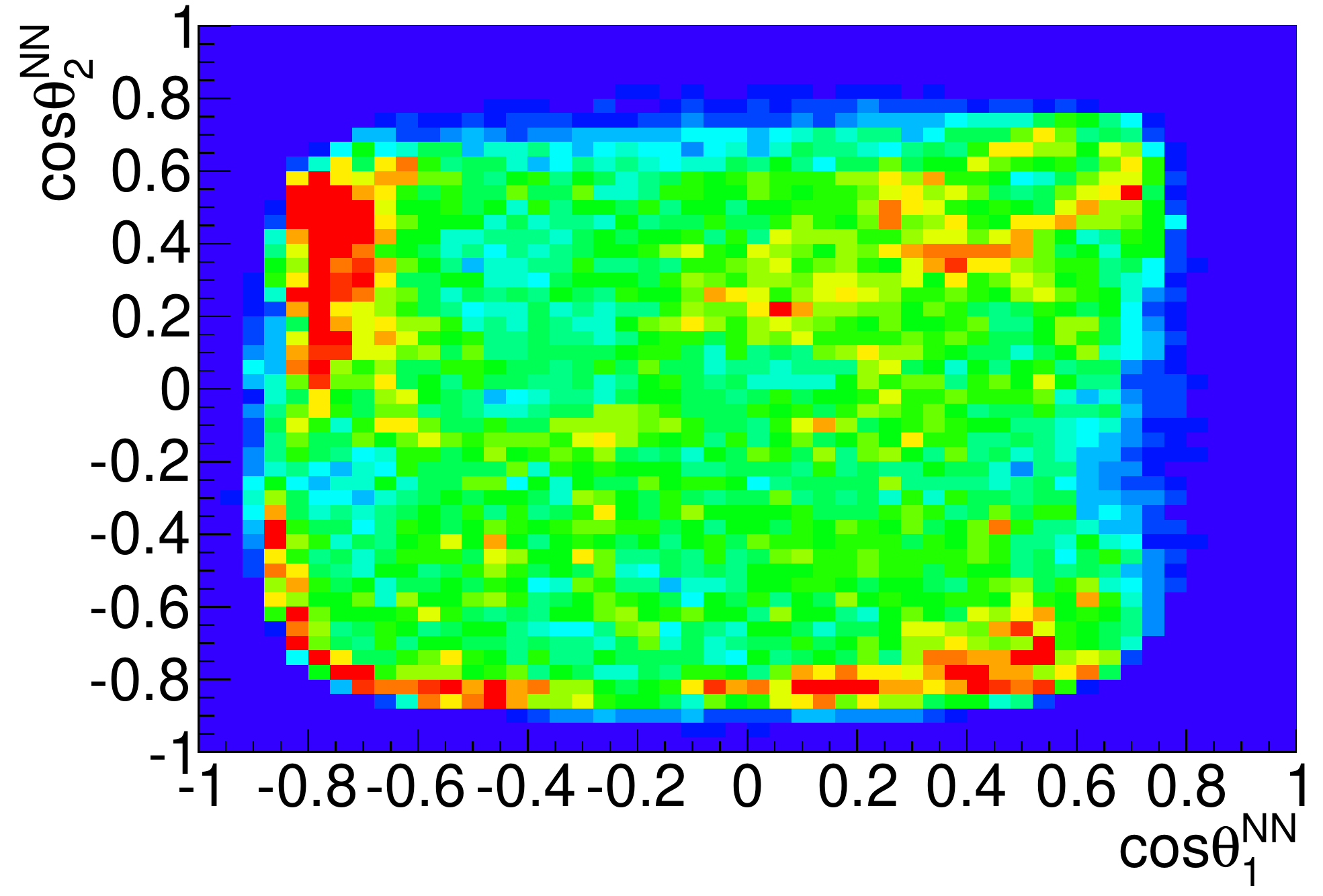}&
\includegraphics[width=.28\textwidth,height=.18\textheight]{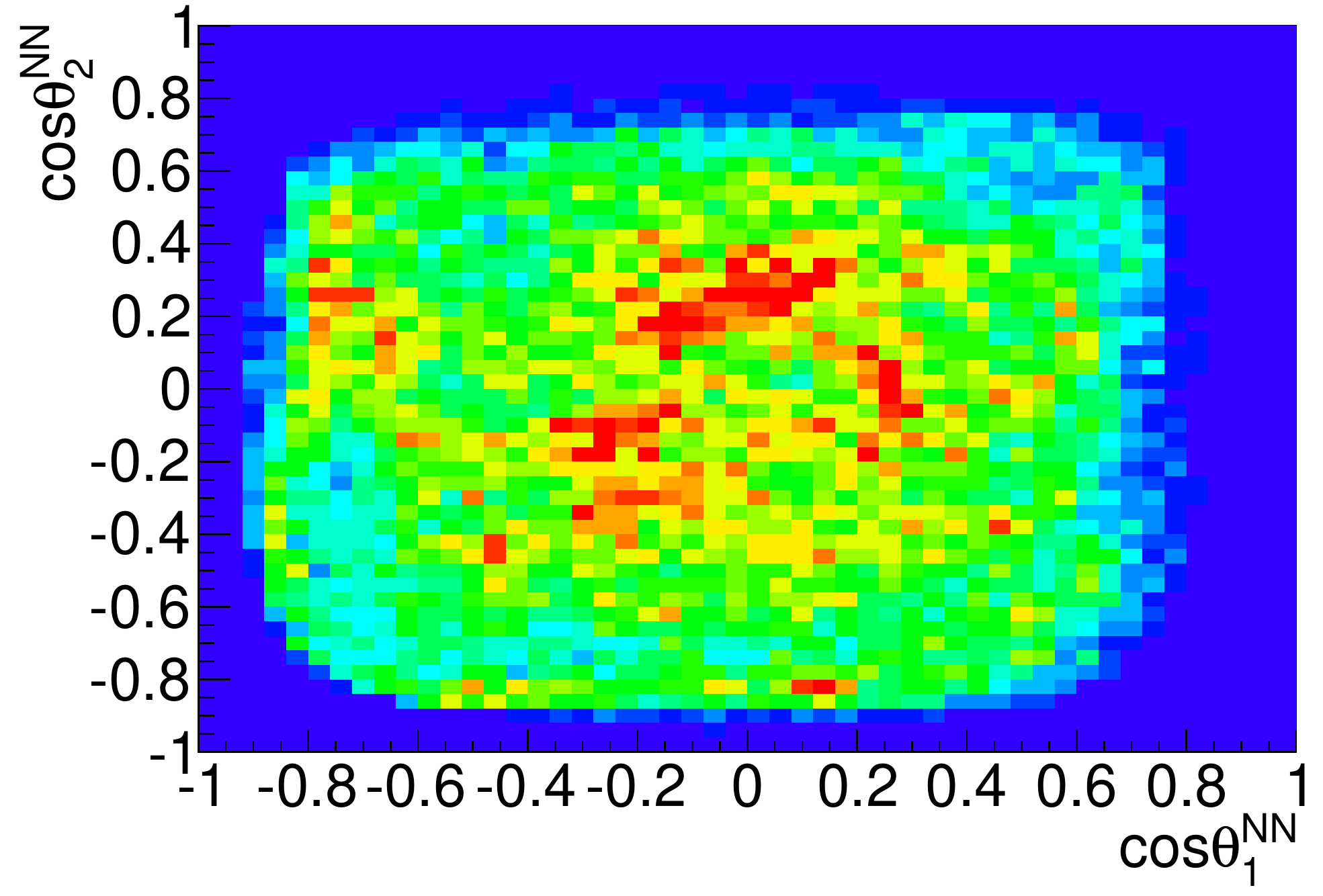}&
\includegraphics[width=.39\textwidth,height=.18\textheight]{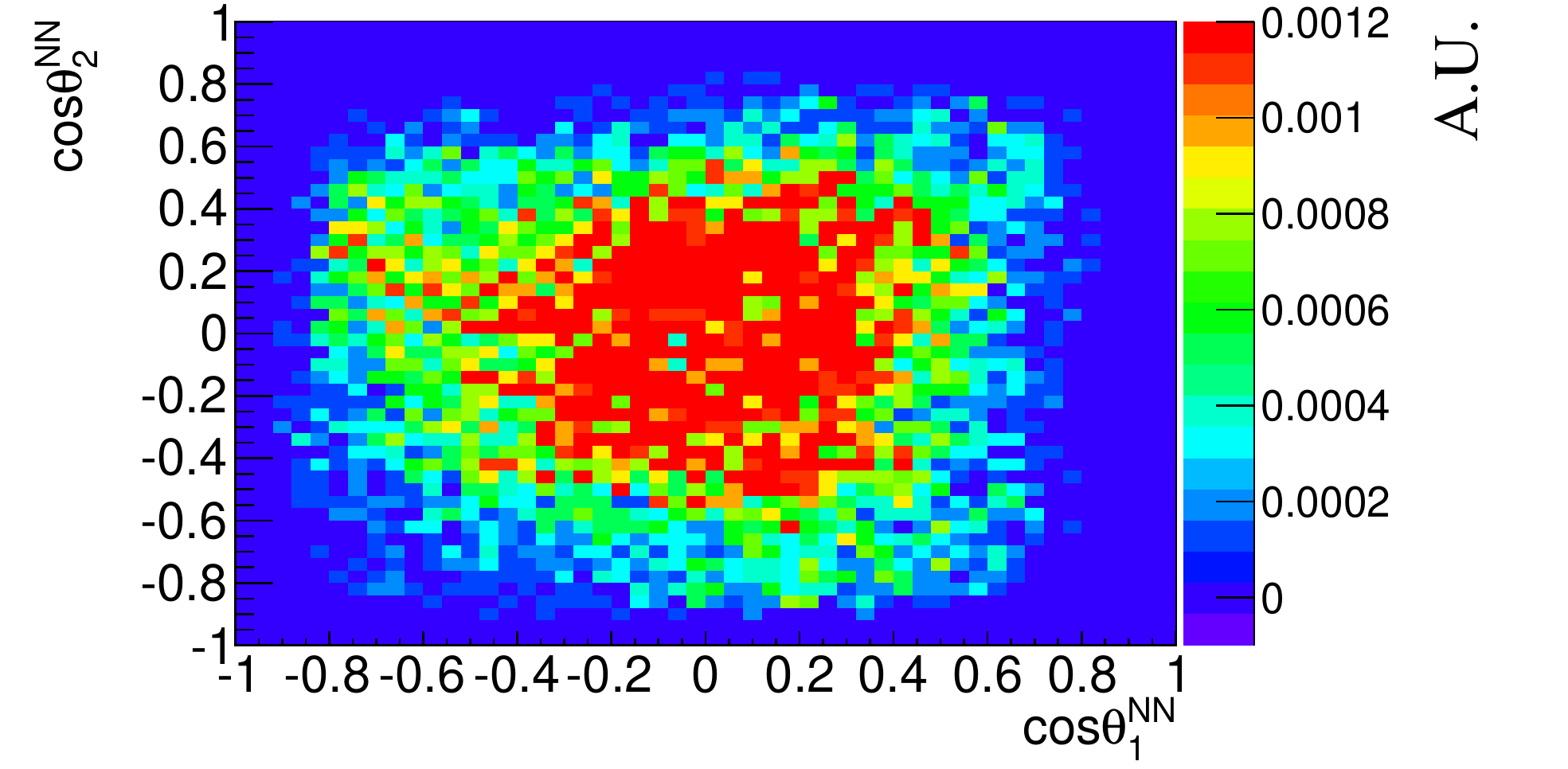}\\
\end{tabular}
\caption{\label{fig:2d_templates} The two dimensional distribution of $\ctsNN_1$ versus $\ctsNN_2$ for the two leptons in each event is shown for each of the three polarization states: $TT$, $TL$, and $LL$, respectively.}

\end{figure*}

Having established templates for each polarization state and a
distribution that is sensitive to different polarization states, 
we fit the two-dimensional $\ctsNN_1$ versus $\ctsNN_2$
distribution in pseudo data to derive each polarization fraction. Five equal-size bins 
are used for each \ctsNN variable ranging from $-1$ to 1. A maximum likelihood fit is performed 
within the RooFit framework~\cite{RooFit}. We combine events with both $W$ bosons transversely-polarized as
``$TT$'' (the sum of $--$, $+-$ and $++$ combinations), events with
one $W$ boson transversely-polarized and one $W$ boson
longitudinally-polarized as ``$TL$'' (the sum of $L-$ and $L+$
combinations), and events with both $W$ bosons
longitudinally-polarized as ``$LL$''. This reduces the free fitting parameters from five to two and 
allows for a better constraint on the $LL$ scattering fraction of interest, under the assumption that
the relative admixture of contributions within $TT$ and $TL$ does not change. The two dimensional distributions for these templates are shown in Fig.~\ref{fig:2d_templates}. Statistical fit uncertainties are determined by randomly
fluctuating data expectations within their Poisson uncertainties and
repeating the fit, and confidence intervals are derived from these toy experiments.  

\begin{figure}[h]

\includegraphics[width=.45\textwidth]{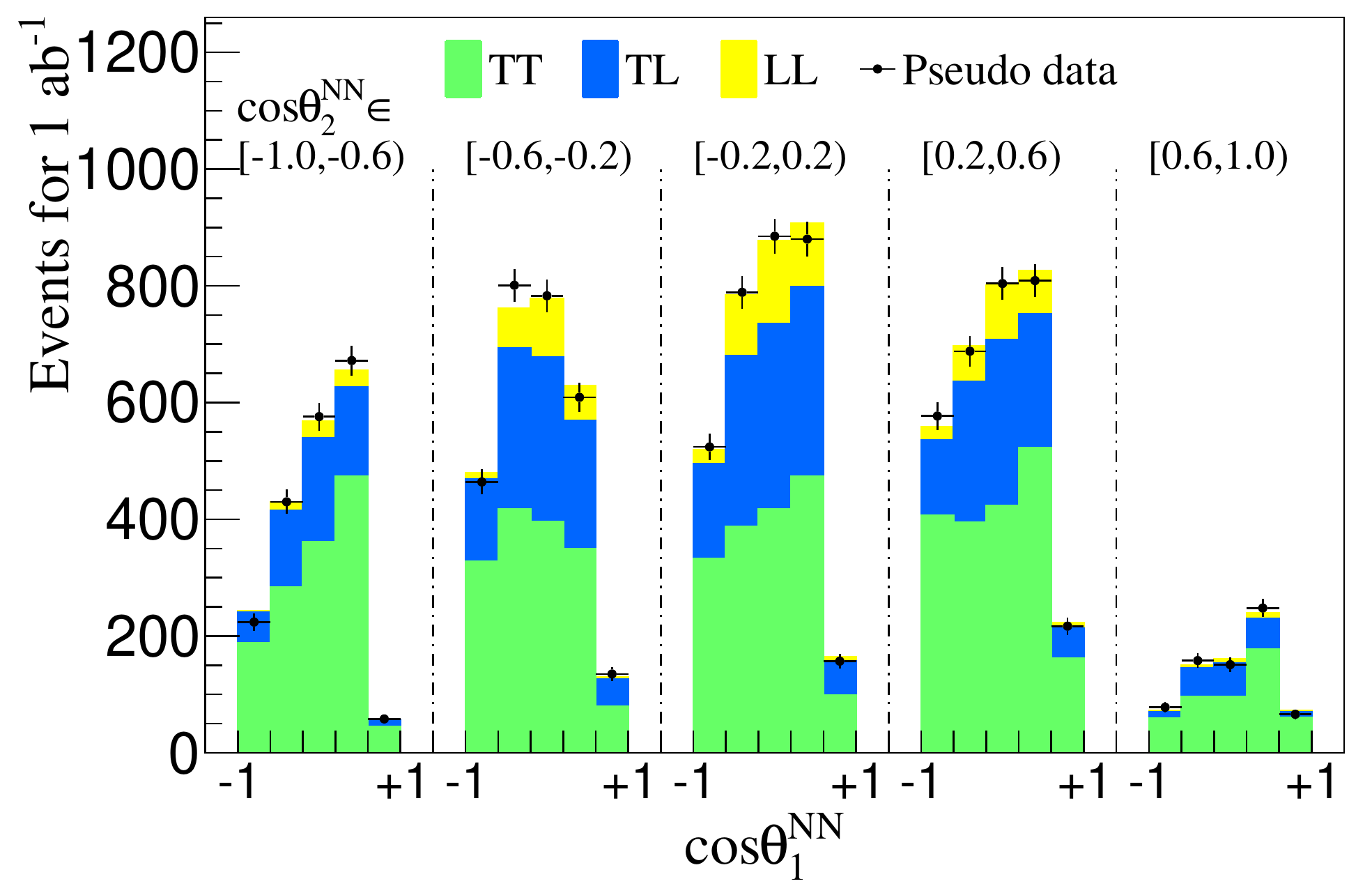}


\caption{\label{fig:fit_example} One example fit where the pseudo data are compared to the sum of 
contributions from $TT$, $TL$ and $LL$ components. There are five groups with five bins inside each group. 
These five groups represent $\ctsNN_2$ from $-1$ to 1 with a step size of 0.4, while five bins inside each group represent $\ctsNN_1$ from $-1$ to 1 with a step size of 0.4. }
\end{figure}

Fits are performed in a range of integrated luminosities from $0.01-3$~ab$^{-1}$. An example fit for 1 ab$^{-1}$ is shown in Fig.~\ref{fig:fit_example} where the pseudo data are compared to the sum of 
contributions from $TT$, $TL$ and $LL$ components. It is found that the $LL$ fraction can be measured with a 68\% confidence limit of 6.7$\pm 1.4\%$ with an ultimate luminosity of 3 ab$^{-1}$. When a similar fit is applied to the $R_{pT}$ variable the precision of 6.7$^{+7.1}_{- 6.7}\%$ is found to be consistent with 0. The regression technique hence greatly enhances the sensitivity to this process.

The above measurements are performed with parton level predictions. While they show encouraging results  it is important 
to also check if this procedure will stand up to experimental reality of finite detector resolution, and the event level selection that
will be required to remove backgrounds from this analysis. To study the effects of event level cuts, we apply additional selection criteria as used by the ATLAS collaboration~\cite{ATLAS_ssWW} to obtain a
tighter fiducial region which is dominated by the contribution from electroweak production of \ssWW events: jet $p_T > 30$ GeV, lepton $p_T > 25$ GeV, missing transverse energy $\slashed{E}_T > 40$ GeV, 
dijet mass $M_{jj} > $500 GeV, and dijet pseudorapidity difference $|\Delta \eta_{jj}| > 2.4 $.

To emulate the response of a typical general-purpose LHC detector,
these events are passed through the application of parton showering in {\sc pythia}\cite{pythia} and
then through the response simulation of the CMS detector implemented
in {\sc delphes}~\cite{delphes}. This detector smearing adds some
degree of realism, but could neglect various effects due to the large
number of overlapping interactions during the high luminosity LHC
runs. Since these effects are often mitigated with specific
reconstruction techniques and need detailed detector modeling, we
leave studies of this nature to dedicated efforts by the experiments.
After detector simulation, parton level quantities are then
approximated by taking the leading two jets clustered with an
anti-$k_t$ algorithm~\cite{antikt} with jet size parameter R=0.5.

Backgrounds to the \ssWW process depend largely on experimental choice, and require detailed simulation. It can be seen in Fig.~\ref{fig:polarization_comparison} that background from the $WZjj$ process (where one of 
the leptons from the $Z$ boson decay is not detected) has a different \ctsNN shape than \ssWW events, but it is likely event level cuts will still need to be applied to reduce this component. Determining the systematics on the background shapes will require significant work from the experiments, and is not treated here. However, as already mentioned the backgrounds in the \ssWW channel are relatively small compared to other channels.

We determine the precision that can be achieved for fractions of $TT$, $TL$, and $LL$ components using three different scenarios increasing in realism: (a) using all  generated events at the parton level; 
(b) using events passing the additional selection criteria used by ATLAS at the parton level; 
(c) using events processed with the {\sc delphes} detector simulation and with reconstructed objects passing additional selection criteria used by \mbox{ATLAS}. 

\begin{figure*}[h]
\includegraphics[width=.9\textwidth]{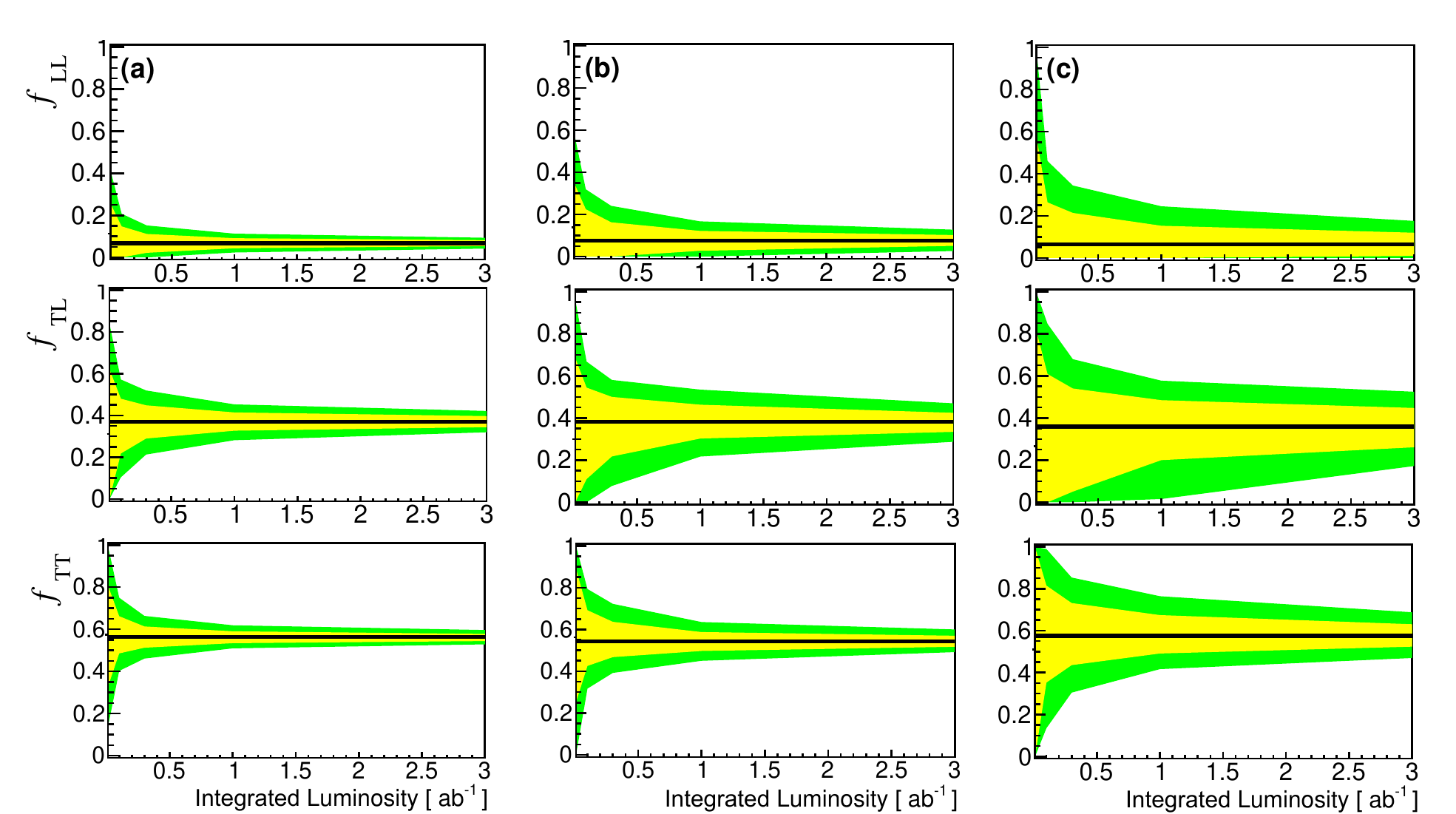}
\caption{ \label{fig:sensitivity} 68\% (yellow) and 95\% (green) expected confidence intervals for the measured $LL$ fraction $f_{LL}$ (top), the $TL$ fraction $f_{TL}$ (middle), and $TT$ fraction $f_{TT}$ (bottom) as a function of the integrated luminosity for, from left to right, scenarios (a),(b), and (c) discussed in the text.}
\end{figure*}

The precision for the three polarization fractions as a function of
the integrated luminosity is presented in Fig.~\ref{fig:sensitivity}.
$TT$ components can be measured with great precision, whereas separating
pure $LL$ scattering from $LT$ scattering is challenging.  In the most
difficult and realistic scenario (c), the cuts and object efficiencies
slightly increase the mean $LL$ fraction to 7.0\% and 68\% confidence
limits are found 7$^{+19}_{-7}$\% (7$^{+5}_{-6}\%$) for an integrated
luminosity of 0.1 (3) ab$^{-1}$.  Equivalent limits from fits to the
$R_{pT}$ variable are found to be 7$^{+29}_{-7}\%$ and
7$^{+9}_{-7}\%$. The fit to the neural network almost doubles the
ultimate precision to anomalous $LL$ fractions in this
scenario. Reaching the same statistical sensitivity using the $R_{pT}$
variable would require approximately 10 ab$^{-1}$, more than three
times the total expected luminosity of the LHC program. We have shown
that large sensitivity gains can be made with NN regression. In
scenario (c) our simple estimate falls short of the $5\sigma$ criteria
for observation of longitudinal VBS, however, new physics beyond the
SM could greatly enhance this fraction~\cite{VLVLBSM}, making
stringent limits or a future observation of this fraction very
important. A comparison of scenarios (a) and (b) to (c) illustrates
potential gains in sensitivity through cut optimization and improved
detector performance. In addition, the authors hope that experiments
can improve on this methodology by training on fully simulated events
or by improving detector performance (e.g. through upgrades) to
enhance sensitivity.

In conclusion, we present a method to determine the $WW$ polarization fractions in
\ssWW events by using a deep machine learning technique.  This method
allows to recover the charged lepton angular distributions in the $W$
boson rest frame from measurable event kinematics.  The
results obtained from this method show greatly enhanced sensitivity over
the example $R_{pT}$ variable.  Cuts to reject
backgrounds as well as finite detector resolutions reduce the sensitivity as
expected, but the method remains a powerful tool for the study of
polarization fractions in VBS events, almost doubling the ultimate precision when compared to $R_{pT}$.

We would like to thank our colleagues Sally Dawson, Frank Paige, and Olivier Mattelaer for their help and guidance while preparing this manuscript.
The contributions from J.S., L.H. and J.Z. are supported by the U.S. DOE Early Career Grant under contract DE-SC0008062. The work of M.-A.P. is supported by U.S. DOE Contract No. DE-SC0012704. We would also like to acknowledge the role of the U.S. ATLAS Scholar program for fostering collaboration between Michigan and Brookhaven National Laboratory.

\clearpage

\providecommand{\href}[2]{#2}\begingroup\raggedright\endgroup

\end{document}